\documentstyle[twocolumn,aps]{revtex}
\begin{document}
\draft

\title{Hexagon Formation as a Compromise Between Squares and Stripes \linebreak in Vertically Vibrated Granular Layers}
\author{Tae-Wook Ko\footnote{Electronic address: kotawo@vortex.kaist.ac.kr\\
 Tel: +82-42-869-2563 \hspace{.2cm} Fax: +82-42-869-2510}, Seong-Ok Jeong, and Hie-Tae Moon}
\address{Department of Physics, Korea Advanced Institute of Science and
Technology, Taejon 305-701, Korea}
\date{\today}
\maketitle

\begin{abstract}
We propose a spatially continuous and temporally discrete model for pattern formation in vertically vibrated granular layers. The grain transfer and the grain mobility transitions are introduced qualitatively, but explicitly. This model reproduces experimentally observed $f/2$ patterns, including squares in high-mobility conditions and stripes in low-mobility conditions. Hexagons are obtained when a high-mobility condition and a low-mobility condition are alternately applied. We discuss this hexagon formation in connection with the formation of squares and stripes. In addition to the experimentally observed localized oscillon structures, this model also exhibits localized hexagonal structures. These localized hexagonal structures grow by nucleating oscillons of equal phase.    
\end{abstract}
\narrowtext

\section{Introduction}
Spatiotemporal patterns are found in numerous physical, chemical, and biological systems, and have drawn much interests in recent years \cite{cross,moon}.
Like fluid systems (Faraday systems) \cite{cross,gollub}, familiar but poorly understood granular systems \cite{jaeger} show various patterns 
including stripes, squares, hexagons and kinks when vertically vibrated \cite{melo94,melo95,umca,das}. 
Novel two-dimensional localized structures, termed 'oscillons', were also observed in these systems \cite{umca,oscil}. Since the pattern formation phenomena in vertically vibrated granular systems were first reported, many models have been proposed to explain the mechanism \cite{bizon,tsim1,tsim2,cerda,shin,roth,jeong1,jeong2,ott1,ott2,crawford}. Molecular dynamics simulations \cite{bizon} quantitatively reproduce every observed patterns except oscillons and kinks, but do not give the essentials of the mechanism. Phenomenological models \cite{tsim1,tsim2,cerda,shin,roth,jeong1,jeong2} explain some of the phenomena by considering the grain transfer with a few qualitative features.  Especially, the model in Ref. \cite{shin} reproduces stripes, squares, and hexagons by using two simple features, i.e., randomizing impacts and inelastic collisions. 
The continuum coupled map model \cite{ott1,ott2} yields a phase diagram similar to that of the experiments with the particular choice of the map and the introduction of a second length scale and explains the universal bifurcation sequence related to the interaction between temporal period doubling and spatial pattern formation.  
The results obtained by using the order parameter equation \cite{crawford} indicate that the localized structures are not specific to granular systems \cite{fineberg99,fineberg20}.   
Although these models give the key to the mechanism, hexagon formation has not been properly treated. The models of Refs. \cite{bizon,tsim2,shin,ott1} exhibit hexagons, but do not elucidate the hexagon forming conditions. Other models reproduce only squares \cite{cerda} or only stripes \cite{roth,jeong1} besides oscillons.

In this paper, by incorporating the grain transfer \cite{kpkim,metcalf,mobility} qualitatively but explicitly, we propose a model that can cover all patterns including squares, stripes, hexagons, and oscillons. Hexagon formation will be discussed in connection with squares and stripes.  

Previously, we proposed a simple model \cite{jeong1,jeong2} which gave some insights into the dynamics of stripes, oscillons, and kinks.  
The following two features were assumed to study stripes and oscillons \cite{jeong1}:
 
(i) A local excitation organizes only through interaction with
 excitations within some distance;

(ii) The effective interaction or flow is approximately determined
 by the difference between the excitations.  

\noindent Additionally, a period-doubling bifurcation nature of the local dynamics was needed to investigate the kinks\cite{melo95,umca,das,jeong2}.  
With these features and a Poincar\'{e} map-like idea \cite{strogatz},  
we simplified the dynamics of the layer and constructed a model which maps the pattern at the time of the layer-plate collision to the pattern at the time of the next collision. 

However, room exists for more generalizations of the model, for it does not show squares or hexagons yet. In this regard, some authors \cite{oscil,tsim1,cerda,metcalf,mobility} argue that mass conservation has an important role in pattern formation and grain mobility contributes to the selection of squares and stripes. In experiments \cite{melo95,oscil}, squares arise at low forcing frequencies, and at those frequencies, the horizontal mobility is large due to the large dilation of the layer and the small dissipation. However, stripes are observed at high frequencies, where the mobility is small due to the small dilation and the large dissipation.  

In Sec. II, we propose a modified model by introducing mass conservation and grain mobility transitions. The results from numerical simulations are presented and discussed in Sec. III.

\section{Model}
In this modified model, we use densities instead of local heights to consider the grain transfer explicitly.
  The density $\sigma_n(\vec{r}~)$ denotes how many grains are piled 
up on a position $\vec{r}$ in a two-dimensional space at the $n$th time step. 
As mentioned above, the $n$th time step corresponds to the $n$th collision between the layer and the plate. 
Note that in the case of $f/2$ patterns, the densities behave as the local heights of the patterns we observe.

This model maps the densities at the $n$th time step to those at the $(n+1)$th time step through following two stages: 

\begin{eqnarray}
The~ first~ stage:~ 
{\sigma^\prime}_{n+1} (\vec{r}~) &=& \sigma_n(\vec{r}~) + \Delta \sigma (\vec{r}~)
,
\label{del_sigma}
\\ \nonumber \\
The~ second~ stage:
\partial_t \sigma(\vec{r}, t) &=& D {\nabla_\bot}^2 \sigma(\vec{r}, t),
\label{diffusion}
\\
\rm{or~ equivalently}, \nonumber \\ 
\bar {\sigma}_{n+1}(\vec{k}~) &=& e^{-Dt_d k^2} \bar {\sigma^\prime}_{n+1} (\vec{k}~), 
\label{fourier}
\end{eqnarray}
where $\Delta \sigma (\vec{r}~)$ is the increment or the decrement in the density at $\vec{r}$, $D$ is a diffusion coefficient, $t_d$ is the effective time duration of diffusion, and ${\nabla_\bot}^2 = \partial^2_x + \partial^2_y$. In Eq. (\ref{fourier}), $\bar {\sigma}_{n}(\vec{k}~)$ denotes the spatial Fourier transform of  ${\sigma}_{n}(\vec{r}~)$. 

The two stages are similar to those of the model proposed in Ref. \cite{cerda}, but here we can control the grain mobility. Also, as our original model \cite{jeong1,jeong2}, this model shares a common spirit with the model of Ref. \cite{ott1,ott2}: these models are spatially continuous, but temporally discrete.

{\it The First Stage.}------ 
Since there is a limit on the distance a grain can travel during the 
time interval between the layer-plate collisions, we take the inside of a circle with radius $R$ centered at $\vec{r}$ in two-dimensional space as an interaction region. Some of the grains at this position $\vec{r}$ will flow into the interaction region ${\mathcal{N}}{(\vec{r}~)} = \left\{\vec{r~}^\prime~|~ 0 < |\vec{r~}^\prime - \vec{r}~| \le R \right\}$. 
 The radius $R$ can be determined by the horizontal velocity, the free flight-time, and the collisions with other grains. For simplicity, we set $R$ to the same constant for all $\vec{r~}$'s. 

The number of grains transferred from one position to another depends largely on the lateral velocity of the grains and the free-flight time during which the lateral transfer of grains occurs \cite{melo95,cerda}.  The lateral velocity is approximately proportional to both the layer-plate collision velocity and the slope of the free surface of the layer \cite{cerda}; the free flight-time is roughly proportional to $1/f$. 

From these facts, we consider the quantities 
 
\begin{eqnarray}
\delta_n (\vec{r} \rightarrow \vec{r~}^\prime) ~~~~~~~~~~~~~~~~~~~~~~~~~~~~~~~~~~~~~~~~~~~~\nonumber \\ 
=  \frac{\alpha}{S({\mathcal{N}}(\vec{r}~))} G\left(\sigma_n (\vec{r}~) - \sigma_n (\vec{r~}^\prime)\right) M_\varepsilon \left(\frac {S({\mathcal{N_<}}(\vec{r}~))} {S({\mathcal{N}}(\vec{r}~))} \right),
\label{delta_n} 
\end{eqnarray}

\begin{equation}
\Sigma_n(\vec{r}~) = \int_{{\mathcal{N_<}}(\vec{r})}d\vec{r~}^\prime~ \delta_n
 (\vec{r} \rightarrow \vec{r~}^\prime), ~~ \vec{r~}^\prime \in {\mathcal{N_<}}(\vec{r}~),
 \label{sn}
\end{equation}
where $ {\mathcal{N_<}}{(\vec{r}~)} = \{\vec{r~}^\prime \in {\mathcal{N}}(\vec{r}~)~|~ {\sigma_n (\vec{r~}^\prime)< \sigma_n (\vec{r}~)}\}$. 
The quantity $\delta_n (\vec{r} \rightarrow \vec{r~}^\prime)$ in Eq. (\ref{delta_n}) is the maximum number of grains that can be transferred from $\vec{r}$ to $\vec{r~}^\prime$ where the density is lower than $\vec{r}$ if there are enough grains at position $\vec{r}$. We call this maximum number the demand from $\vec{r~}^\prime$ to $\vec{r}$. The actual number transferred is limited by the density at $\vec{r}$. For this purpose, $\Sigma_n(\vec{r}~)$, the total demand to $\vec{r}$, is given in Eq. (\ref{sn}). 
In this model, the parameter $\alpha$ acts like the layer-plate collision velocity (or the relative collision acceleration) which determines the onset of patterns \cite{melo95,cerda}.
The dependence of the transferred number on the slope of the free surface is phenomenologically represented by the function $G$, which depends only on the difference between the densities at the two positions. Due to the avalanche-type flow \cite{jaeger,aval}, the number of grains transferred is naturally expected to increase faster than linearly as the difference increases, so we choose a slope-increasing function for $G$ [$G^{\prime \prime} > 0$; see Fig. 1 (a)]. As in our previous model \cite{jeong1,jeong2}, this slope-increasing property of the function $G$ makes the hysteresis more evident and strong: As patterns grow from the small fluctuations in the densities, the density differences increase more, and the effect of this coupling function becomes more important. Therefore, this effect makes the quantities in Eqs. (\ref{delta_n}) and (\ref{sn}) large enough to support the excitations even if $\alpha$ is lowered below the critical values where the transitions from flat to patterns occur (see Fig. 4 to be discussed). 
The function $M_\varepsilon$ controlling the  mobility of the grains at a position $\vec{r}$ is a function of the status of the position $\vec{r}$ in the interaction region  ${\mathcal{N}}(\vec{r}~)$. $\varepsilon$ is a control parameter. 
The status of a position $\vec{r}$ is   
given by the quantity ${S({\mathcal{N_<}}(\vec{r}~))}/{S({\mathcal{N}}(\vec{r}~))}$, where  
$S({\mathcal{N_<}}(\vec{r}~))$ is the area with densities lower than $\sigma_n(\vec{r}~)$  in the interaction region  ${\mathcal{N}}(\vec{r})$, and $S({\mathcal{N}}(\vec{r}))$ is the total area of the interaction region. Thus, positions with relatively higher densities have higher status. For example, a peak in the layer corresponds to the highest status and a valley to the lowest one. Experimentally, there are qualitative changes in the overall mobility as the forcing frequency changes.  The mobility is high at low frequencies while at high frequencies the grains are less mobile.
Without $M_\varepsilon$, in this model, the grains at positions with high status are more apt to move than those at positions with low status due to the large values of the quantities $\delta_n (\vec{r} \rightarrow \vec{r~}^\prime)$ and $\Sigma_n(\vec{r}~)$. We can control the relative mobilities between these positions by giving different $M_\varepsilon$ values to them according to their status. If we give high $M_\varepsilon$ values to positions with low status [see Fig. 1 (b), $\varepsilon < 0$], the mobilities at these positions increase. In such a case, most of the grains are apt to move. Contrary to this case, if we give low $M_\varepsilon$ values to positions with low status [see Fig. 1 (b), $\varepsilon > 0$], the grains at these positions are even more immobile. 
The parameter $\varepsilon$ in our model roughly plays a role analogous to that of the forcing frequency in the experiments. 
The denominator $S({\mathcal{N}}(\vec{r}~))$ in Eq. (\ref{delta_n}) is introduced to diminish the $\alpha$ value dependence on the radius of the interaction region ${\mathcal{N}}(\vec{r})$.

Since there are limited number of grains at the position $\vec{r}$ and we want to conserve the mass, satisfying equally the demands from $\vec{r~}^\prime$, we distribute the limited number of grains to ${\mathcal{N_<}}(\vec{r}~)$ in two possible ways:  

Case 1: If the total demand $\Sigma_n(\vec{r}~)$ is less than the density at $\vec{r}$,
the outflow from  $\vec{r}$ to $\vec{r~}^\prime \in {\mathcal{N_<}}(\vec{r}~)$ is set to the demand $\delta_n (\vec{r} \rightarrow \vec{r~}^\prime)$.

Case 2: If the total demand $\Sigma_n(\vec{r}~)$ is greater than the density at $\vec{r}$, the total outflow from $\vec{r}$ is set to the density at $\vec{r}$, and the outflow is 
distributed to ${\mathcal{N_<}}(\vec{r}~)$ in proportion to the demand $\delta_n (\vec{r} \rightarrow \vec{r~}^\prime)$.  

\noindent Note that grains are more actively transferred in Case 2 than in Case 1, for in such a case, all the grains at those positions leave their positions. 

In this manner, we can calculate the inflow and the outflow of the grains for each position at this first stage.   

{\it The Second Stage.}------
 In the second stage, to include the diffusion effect due to the collisions between grains and between the layer and the plate \cite{cerda,shin}, we apply Eq. (\ref{diffusion}) which automatically conserves mass. 
This process relaxes the high-density gradient occurring within the size of the characteristic length scale $\lambda_0 = 2R$. Since the net diffusion effect is determined by the parameter $D t_d$, we choose $D t_d \approx (\frac {2 \pi}{\lambda_0})^{-2}$ such that the Fourier components of the density distribution with wave vectors greater than $k_0 = \frac {2 \pi}{\lambda_0}$ are sufficiently suppressed. 

\section{Numerical Results and Discussion}

For the numerical simulations, we choose the functions (see Fig. 1) 

\begin{eqnarray}
G(x) &=& x + c x^2,~ c > 0, x > 0   , 
\label{g_x}\\
M_\varepsilon(x)&=& x^{\varepsilon},~ 0 < x \le 1 .
\label{m_x}
\end{eqnarray}

\noindent These functions are the simplest ones with the required properties.  
$G(x)$ in Eq. (\ref{g_x}) is a monotonically increasing function 
 with a slope-increasing property ($G^{\prime \prime} > 0$). $M_\varepsilon$ in Eq. (\ref{m_x})
 is  monotonically increasing for $\varepsilon > 0$, but  monotonically decreasing for $\varepsilon < 0$. 
Other functions with the same properties do not give qualitatively different results. We fix $c = 0.7$ in the following simulations for the stability of oscillon structures. Higher (lower) $c$ values give larger (smaller) hysteresis.  

We performed numerical simulations on a $128 \times 128$ grid with spatial extension $L=128$ and periodic boundary conditions.  
 Initially, the densities are distributed around the mean value $\sigma_0 = 10$ with small uniform random deviations within $[-10^{-3}, 10^{-3}]$. 

Square and stripe patterns obtained using this model are shown in Figs. 2 and 3.
These patterns oscillate with $f/2$ as in experiments \cite{melo94,melo95,umca}; that is, they return to the same configuration after 2 periods of the forcing (in this case, one time step corresponds to one period).
  Figure 4 shows the stability region for each pattern as a function of $\alpha$ and $\varepsilon$.  
There are two transition regions between squares and stripes. At lower $\alpha$'s, the transition from squares to stripes occurs as the  parameter $\varepsilon$ increases from negative to positive values, and for positive $\varepsilon$'s, the transition from stripes to squares occurs as $\alpha$ increases. Both of these transitions correspond to qualitative changes in the grain mobility. As expected, squares arise for $\varepsilon < 0$ or sufficiently large $\alpha$, where the quantities in Eqs. (\ref{delta_n}) and (\ref{sn}) are large enough that grains are actively transferred: All the grains at positions of more than $85\%$ of the entire space leave their positions according to Case 2 at the first stage of our model. On the contrary, when stripes appear, the percentage decreases below $80\%$.       
For the same reason, greater $c$ values for $G(x)$ in Eq. (\ref{g_x}) lowers the transition points from stripes to squares for positive $\varepsilon$ values.

Figure 3 also reflects the grain mobility transitions. As pointed out in Ref. \cite{shin}, the densities are sharply clustered in the square patterns (the high-mobility situation) and more diffuse in the stripe patterns (the low-mobility situation); the peaks in the square pattern are higher than those in the stripe patterns (see Fig. 3). 

In the hysteretic region, this model also exhibits period-2 oscillons and the oscillon structures such as dipoles, triangular tetramers, and chains (not shown), which are observed experimentally \cite{oscil,umca}.  When $\alpha$ is increased above the upper stability limit, the structures grow by nucleation of oscillons ($\varepsilon < 0$ in this model, low $f$ in experiments) or by elongation of oscillons and nucleation of waves ($\varepsilon > 0$ in this model,  $f > 30$ in experiments) \cite{umca,oscil,jeong1}. 

Now, we consider the hexagon formation. Hexagons are also period-2 patterns like squares and stripes, but there is a difference. Two distinct phases -- a set of isolated peaks on a triangular lattice and a hexagonal cell -- appear in alternate cycles. 
Experimentally, hexagons arise when effective two-frequency forcing is applied internally or externally \cite{melo95,oscil,umca}. In this situation, the layer has two different free-flight times and, consequently, collides with the plate alternately with two different colliding conditions. 
In the absence of any clear identification of the colliding conditions \cite{melo95,ott2}, from the two phases of hexagons, we conjecture that one is a high-mobility condition, and the other is a low-mobility condition. To determine the validity of our conjecture, we apply a high-mobility condition (a square-forming condition) in one step and in the next step a low-mobility condition (a stripe-forming condition) with the same radius.  This yields hexagonal patterns as shown in Figs. 5 and 6.

As in experimental results \cite{melo95,umca}, hexagonal patterns obtained in this model show two different phases - a peak phase and a cellular phase (Fig. 5). Note that the peak phase has localized density peaks like squares and always appears after applying the high-mobility condition [see Fig. 3(a), Fig. 6(a), Fig. 5(a)]. Similarly, note also that the cellular phase has ridges like stripes and always arises after applying the low-mobility condition [see Fig. 3(b), Fig. 6(b), Fig. 5(b)].  From these facts, we can regard the hexagon formation as a compromise between squares and stripes.  Our results are consistent with the occurrence of hexagons in the transition region between squares and stripes in Fig. 2 of Ref. \cite{shin}.
 However, the hexagon formation described by the order parameter equation \cite{tsim2} is different from ours. In our model, grain transfer is crucial for hexagon formation, but in the case of Ref. \cite{tsim2} grain transfer is treated irrelevant to hexagon formation.   

Since the hexagons in this model are formed directly from the flat state and since the transitions are hysteretic, we have parameter regions with coexistence of hexagons and flat. In that region, we obtain stable localized hexagonal structures as shown in Fig. 7. These structures may be regarded as oscillon structures. Actually, if the parameter $\alpha$ is raised above the upper stability limit, these hexagonal structures grow into a global hexagonal pattern by nucleating oscillons of equal phase at the hexagonal lattice sites (see Fig. 8) as the oscillon structures at low forcing frequency grow through square-like structures into squares and at high frequency through stripe-like structures into stripes \cite{umca,oscil}.     
Although such localized structures have not been reported in granular systems, similar structures have been found in other systems \cite{fineberg20,conv,visco,coullet1,coullet2}. Especially, the localized hexagonal oscillon structures found in the two-frequency forcing experiments with Newtonian fluids \cite{fineberg20} seem to be closely related to the structures found in our model. 
The growing phenomena reported in Refs. \cite{fineberg20,coullet1,coullet2} are also similar to the one described above. In contrast, the hexagonal patch in Ref. \cite{conv} grows by a six-faceted, moving front, and the three independent roll states in superposition give the hexagonal structure.

\section{conclusion}
In summary, considering the grain transfer qualitatively but explicitly, we have constructed a model for pattern formation in vertically vibrated granular layers. Our model maps the density distribution at the time of the layer-plate collision to that at the time of the next collision in a continuous two-dimensional space. This model exhibits squares in high-mobility conditions and stripes in low-mobility conditions. 
To elucidate the hexagon forming situation, we simulated period doubling situation by applying a high-mobility condition (a square-forming condition) and a low-mobility condition (a stripe-forming condition) alternately and observed hexagons. Our simulations show that the peak phase of hexagons appears after the high-mobility condition and the cellular phase after the low-mobility condition. The peak phase has localized density peaks like squares, and the cellular phase has ridges like stripes. From these facts, we conclude that in vertically vibrated granular layers hexagon formation is a compromise between squares and stripes. 
Finally, the model presented here also reproduces the experimentally observed oscillon structures and their dynamics. In addition, we observed localized hexagonal structures and their growth by nucleation.  Our model may be useful for the studies of multifrequency forcing experiments \cite{umca,fineberg99,fineberg20,pak} and other pattern forming systems \cite{cross,salt_f}.

\section*{acknowledgments} 
We thank  Hwa-Kyun Park for useful discussions.
This work was supported in part by the  the Interdisciplinary Research Program
(Grant No. 1999-2-112-002-5) of the Korea Science and Engineering Foundation and in part by the BK21 program of the Korea Ministry of Education. We also appreciate the support of the Korea Research Foundation (Grant KRF-2000-DP0097).

\newpage
{\large \bf Figure Captions}
\begin{description}

\item{Fig. 1.} Functions in Eq. (\ref{delta_n}). (a) $G(x)$ is a function of the density difference between two positions. (b) $M_\varepsilon$ is a function of the status (see text for definition). the parameter $\varepsilon$ controls the overall grain mobility.  

\item{Fig. 2.} Shadow graph of the patterns obtained numerically in our model. (White denotes high density.) (a) Period-2 square pattern at $\alpha = 3, \varepsilon = -3, R = 6$, and $Dt_d = 3.65$. (b) Period-2 stripe pattern at $\alpha = 11, \varepsilon = +3, R = 6$, and $Dt_d = 3.65$.

\item{Fig. 3.} Three-dimensional perspective of the densities in (a) the square and (b) the stripe patterns of Fig. 2. 

\item{Fig. 4.} Phase diagram of our model as a function of $\alpha$ and $\varepsilon$ with $R = 6$ and $Dt_d = 3.65$. 
 Closed (open) squares denote transitions from flats (patterns); closed triangles and circles indicate the transition points. The numbers near symbols are the corresponding $\alpha$ values. 
	  
\item{Fig. 5.} Hexagonal pattern obtained by alternately applying a high-mobility condition and a low-mobility condition. (a) Peak phase $(n = odd)$ after applying the high-mobility condition $(\alpha_1 = 5, \varepsilon_1 = -3)$ and (b) cellular phase $(n = even)$ after applying the low-mobility condition $(\alpha_2 = 5, \varepsilon_2 = +3)$. $R_1 = R_2 = 6$ and $(Dt_d)_1 = (Dt_d)_2 = 3.65$.

\item{Fig. 6.} Three-dimensional perspective of the densities in the hexagonal pattern of Fig. 5 in (a) the peak phase and (b) the cellular phase.   

\item{Fig. 7.} Localized hexagonal structure in the hysteresis region 
obtained by applying the same conditions as in Fig. 5, except for $\alpha_1 = \alpha_2 = 1.1$. 

\item{Fig. 8.} Growth of hexagonal structure by nucleation. The $\alpha$ values
are raised from $1.1$ in Fig. 7 to $1.5$. 

\end{description}
\end{document}